\documentclass{article}
\usepackage{graphicx} 
\usepackage{authblk}
\usepackage{float}
\usepackage{amsmath}
\usepackage{amssymb}
\usepackage{caption}
\usepackage{subcaption}

\usepackage{lipsum}

\title{Deep Koopman Operator-based degradation modelling}
\author[1]{Sergei Garmaev}
\author[1]{Olga Fink}
\affil[1]{Intelligent Maintenance and Operations Systems Laboratory, EPFL}

\date{June 2023}

\begin{document}

\maketitle

\begin{abstract}
    Reliable health indicators of industrial systems that can accurately represent the evolution of the true health conditions is of paramount importance for condition monitoring, fault detection and  reliable prediction of the remaining useful lifetime. However, constructing such indicators is a non-trivial task and typically requires domain specific knowledge. With the current trend of increasing complexity of industrial systems, the construction and monitoring of health indicators becomes even more challenging. Given that health indicators are commonly employed  to predict the end of life, a crucial criterion for reliable health indicators is their capability to discern a degradation trend. However, trending can pose challenges due to the variability of operating conditions. An optimal transformation of health indicators would therefore be one that converts degradation dynamics into a coordinate system where degradation trends exhibit linearity.  Koopman theory framework is well-suited to address these challenges. In this work,  we demonstrate  the successful extension of the previously proposed Deep Koopman Operator approach to learn the dynamics of industrial systems by transforming them into linearized coordinate systems, resulting in a latent representation that provides sufficient information for estimating the system's remaining useful life. Additionally, we propose a novel Koopman-Inspired Degradation Model for degradation modelling of dynamical systems with control. The proposed approach effectively disentangles the  impact of degradation and imposed control on the latent dynamics. The algorithm consistently outperforms in predicting the remaining useful life of CNC milling machine cutters and Li-ion batteries, whether operated under constant and varying current loads. Furthermore, we highlight the utility of  learned Koopman-inspired degradation operators analyzing the influence of imposed control on the system's health state.
\end{abstract}

\section{Introduction}

Constructing reliable health indicators is crucial for predictive maintenance. The construction of such indicators is a non-trivial task and requires domain specific knowledge. We highlight that the current trend of increasing real time monitoring data and computational resources availability provides both opportunities and challenges. The modern data-driven approaches \cite{pimenov2023artificial} deliver the means to tackle increased data availability and allow forward prediction of the remaining useful life (RUL), as well as construction of health indicators. While forward RUL prediction requires extensive sets of run-to-failure trajectories for training, data-driven construction of robust system's health indicators can be more promising approach. Formulating precise requirements for construction of health indicators is a challenging task, however, three common characteristics can be highlighted: trendability, prognosability, and monotonicity \cite{coble2009fusing}.

Estimation of the system's state of health can be performed by utilizing the knowledge of underlying physics and mechanisms governing the degradation process \cite{daigle2012model, daigle2016end}.  Such degradation models provide  high interpretability and generalizability, but their usefulness is restricted due to  limited applicability and reliance on simplifications and assumptions. Another approach  is to leverage  modern deep learning techniques  to extract health indicators and estimate RUL using  these indicators \cite{xu2021machine}. Alternatively, deep learning-based methods can be employed for forward RUL prediction, which involves a direct mapping from sensor readings to RUL. However, it is important to note that forward RUL prediction necessitates  a substantial amount of  data for supervision \cite{li2022end, li2019deep, rauf2022machine}. Incorporating physics and prior system knowledge into data-driven model training has been shown to enhance  predictions \cite{xu2022physics, jiang2022model} and reduce data requirements \cite{chao2022fusing}. However, typically in industrial systems , preventive maintenance is widely practiced to mitigate the potential consequences of failures. This practice, while beneficial for risk reduction, often leads to a  limited availability of ground truth run-to-failure data. Furthermore, the challenge  of constructing universal trendable representations of system's health state remains unresolved. A vital requirement for reliable health indicators is their ability to discern a degradation trends accurately. Hence, an optimal transformation of health indicators would involve converting degradation dynamics into a coordinate system where degradation trends exhibit linearity. In this context, we propose to utilize the capabilities offered by Koopman's operator theory.

The Koopman's operator theory \cite{koopman1931hamiltonian} provides  a flexible framework for modelling nonlinear dynamical systems. This theory offers a way  to discover intrinsic coordinate systems where nonlinear dynamics can be expressed in a linear form. Acquiring linear representations of highly nonlinear systems is particularly valuable  for controlling and predicting their dynamic behaviour. Initially, the deep learning capabilities were employed in \cite{Lusch2017} to approximate the eigenfunctions of Koopman operators. The work of Yeung et al.\cite{yeung2019learning} introduced deep dynamic mode decomposition, which outperformed the existing extended dynamic mode decomposition by an order of magnitude in long-term forecasting  \cite{williams2015data}. Another extension using deep learning, proposed in \cite{li2021deep}, addresses  fine-scale dynamics by downscaling a learned coarse model. Several recent works, such as \cite{han2020deep, shi2022deep, ping2021deep} have demonstrated the application of deep learning-based Koopman approaches for optimal control of non-linear dynamical systems. Its probabilistic extension within reinforcement learning framework, as demonstrated in \cite{han2021desko}, has shown superior performance in terms of modelling and control, as well as increased robustness against  large disturbances . However, the proposed setups in these approaches imply a linear mapping of control to Koopman invariant subspace, which limits the choice of the observables. Moreover, these setups may be not applicable for determining the slow dynamics of degradation since the Koopman operator is independent of the imposed control.

This work aims to develop a flexible approach for constructing  health state representation of dynamical systems. To the best of our knowledge, there have been no prior works that attempt to learn hidden health parameters in unsupervised manner using Koopman operator framework. Our objective is to predict RUL of dynamical systems using the latent state representation obtained from the Deep Koopman Operators (DKO). Additionally, we  propose a novel algorithm called Koopman-Inspired Degradataion Model (KIDM) to learn the dynamics of dynamical systems with control and predict their RUL. This approach enables us to leverage the influence of control on the latent state representation and learn the degradataion of hidden parameters that determine the system's dynamics.

In this work we will demonstrate  the applicability of DKO in learning the dynamics of industrial systems and utilizing the learned observables space for accurate  prediction of the system's RUL. We will also demonstrate  that the KIDM is capable of generating informative observables to determine the degradation trend. Both DKO and KIDM exhibit robustness even when only a single ground truth run-to-failure trajectory is available for unseen cases, making them less dependent on extensive run-to-failure data. Additionally, we will conduct ablation studies to evaluate the significance of the Koopman operator components  in the proposed algorithms. The results will reveal that incorporating a loss term for multiple future steps prediction improves the robustness of DKO and KIDM,  leading to reduced  RUL prediction errors. Furthermore, we will demonstrate that forward RUL prediction model struggles  to determine the degradation trend due to the lack of ground truth run-to-failure data. We will further show that contrary to other  ablated models , the KIDM algorithm exhibits a superior  performance in terms of RUL prediction. Additionally, we will conduct an analysis of the extrapolation capabilities of KIDM and show how the learned Koopman-inspired operators can be further examined to assess the influence of imposed control on the system's degradation process.

The remainder of the paper is organized as follows. In the first part of second chapter we briefly outline key concepts of Koopman theory and describe previously proposed Deep Koopman Operator (DKO) framework. Second part describes the newly proposed Koopman-inspired degradation model (KIDM). We then report in the third chapter case studies used to evaluate the performance of DKO and KIDM. Chapters four and five present the results of our studies and conclude the results respectively.

\section{Background on Koopman Operator Theory and Deep Koopman Operator}
In this chapter, we first present  a fundamental  background on the Koopman operator theory.  Subsequently, we provide a concise overview of the original Deep Koopman Operator approach.

The Koopman operator theory was introduced by Bernard Koopman in the early 1930s \cite{koopman1931hamiltonian} . It continues to be an active field of research, with recent developments focused on the deep Koopman Operator approach \cite{brunton2022modern}. The Koopman operator provides a fundamental mathematical framework for analyzing complex dynamical systems. This framework  particularly valuable for studying nonlinear dynamical systems, as it allows for a shift from specific governing equations to a description of the dynamics in the system's state space. Instead of directly considering the finite-dimensional states of the system, which can be highly complex and nonlinear, the Koopman operator approach maps the system's state into an infinite-dimensional space. In this space, the dynamics of the system is described by a linear operator known as the Koopman operator.

A dynamical system can generally be described by the equation:
\begin{equation}
    \dot{x}(t) = f (x(t)),
\end{equation}
where $x \in \mathbb{R}^n$ represents the system state, and $f$ is a vector field that characterizes the system dynamics \cite{brunton2022modern}. In the general case, $f$ is nonlinear.

The Koopman operator operates in an infinite-dimensional space, where the basis functions of this space are typically selected as observables of the system state. Observables refer to quantities that can be measured or observed \cite{bevanda2021koopman}. We refer to the space on which the Koopman operator acts as "observables space" or "observables".

The observables,  denoted as  $y = \phi (x)$ now need to satisfy the equation:
\begin{equation}
    \dot{y} = K y,
\end{equation}
where $K$ is a linear operator that fully determines the system dynamics.

The process of discovering of observables space can be a challenging task. It is possible that a finite-dimensional observables space does not exist for a given system. In practice, the aim is to find a finite-dimensional approximation of the observables. Modern deep learning methods have shown to be valuable tools in addressing this challenge.

\begin{figure}[ht]
    \centering
    \includegraphics[width=0.7\textwidth]{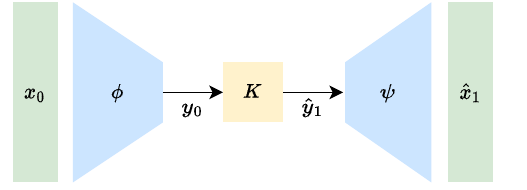}
    \caption{Deep Koopman Operator architecture.}
    \label{fig:dko}
\end{figure}

To leverage the capabilities of deep learning, an extension to the Koopman operator framework was proposed in \cite{Lusch2017}, known as the Deep Koopman Operator (DKO). The DKO  enables the learning of non-linear intrinsic coordinates that linearize the system's dynamics. The DKO architecture is built  upon  a modified auto-encoder structure. The encoder network  approximates the mapping to observables space, while the decoder network performs the inverse mapping. Simultaneously, a learnable linear operator is trained within the observables space to capture the linear dynamics explicitly. Fig. \ref{fig:dko} illustrates  a schematic  representation of the DKO algorithm.

To learn an invertible mapping from the state to the observables space, the state reconstruction loss, defined as $\mathcal{L}_{rec} = \| x_t - \psi(\phi(x_t))\|$, is minimized. The loss term aims to learn the reconstruction of the system state $x_t$ at the time step $t$. To capture linear dynamics within the observables space, the linear dynamics loss, that is defined as $\mathcal{L}_{lin} = \| \phi(x_{t+m}) - \mathbf{K}^m \phi(x_t) \|$, is employed. This loss helps in learning linear dynamics in the observables space by applying the learnable Koopman operator over multiple time steps. Finally, the future state prediction loss, represented by $\mathcal{L}_{pred} = \| x_{t+m} - \psi(\mathbf{K}^m \phi(x_k)) \|$ is introduced  to minimize the prediction error of future states  and facilitate the learning of  dynamics in the initial state space. The overall training objective combines the aforementioned losses as $\mathcal{L} = \mathcal{L}_{rec} + \mathcal{L}_{lin} + \mathcal{L}_{pred}$.





\section{Koopman-Inspired Degradation Model}

Many dynamical systems are subject to external control, which categorizes them  as dynamical systems with control. Dynamical systems with control refer to systems that describe the behavior of a physical process or a mathematical model, taking into account  the influence of an external control input.
Dynamical systems with control find extensive applications in various domains such as  engineering, physics, and other fields. They are particularly useful for designing and optimizing complex systems, where the control input plays a crucial role in shaping the system's behavior and achieving desired outcomes.

Dynamical systems with control are typically described by their dynamics equations:
\begin{equation}
    \dot{x} = f(x, u),
\end{equation}
where $x$ represents  the state of the system, $u$ represents  the control inputs, and the function $f$ captures  the system dynamics.

In practical  applications, it is often more convenient to express  the dynamics in discrete-time form:
\begin{equation}
    x_{t+1} = f(x_t, u_t),
\end{equation}
where $t$ represents  the discrete time step.

The original formulation of the Deep Koopman algorithm \cite{Lusch2017} does not incorporate  control inputs, which  limits its applicability to a narrower range systems. However, in many industrial systems, control inputs  directly  impact  the degradation process, thereby  affecting the health indicators and the RUL of assets. Therefore, considering control inputs is crucial for accurately modeling and predicting the behavior of such systems. Previous works \cite{han2020deep, shi2022deep, ping2021deep, han2021desko, morton2019deep} on incorporating control inputs into DKO frameworks typically employ an architecture where the control vector $u_t$ at time $t$ is multiplied by a linear matrix and added to THE resulting observables. Consequently, the state at the time $t+1$ is defined as:

\begin{equation}
    \hat{x}_{t+1} = \psi( K \phi(x_t) + B u_t ).
\end{equation}

\begin{figure}[ht]
    \centering
    \begin{subfigure}[b]{0.5\textwidth}
        \centering
        \includegraphics[width=\textwidth]{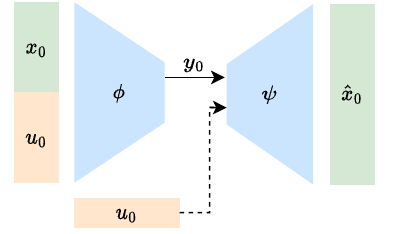}
        \caption{ }
    \end{subfigure}
    \begin{subfigure}[b]{0.95\textwidth}
        \centering
        \includegraphics[width=\textwidth]{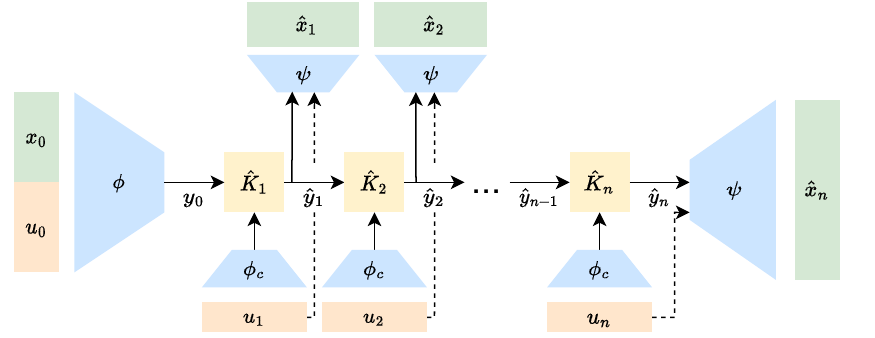}
        \caption{ }
    \end{subfigure}
    \caption{The architecture of Deep Koopman operator with control: (a) prediction pipeline; (b) reconstruction pipeline.}
    \label{fig:kidm}
\end{figure}

However, this formulation implies that control inputs should be linearly  mapped to the observables space, potentially limiting the choice of observables. In this work, we propose  the Koopman-Inspired Degradation Model (KIDM) as an extension to the Deep Koopman operator algorithm.  KIDM enables  the inclusion of control inputs while preserving  information about hidden system health indicators. The proposed architecture follows the encoder-decoder paradigm, where the system dynamics are lifted into a latent state space. However, instead of learning the inverse of observables, the decoder takes both the latent state representation and the control vector as inputs to either reconstruct the current state or predict the future state. This design allows the encoder to extract information about the system's response to a given control input, thereby  enabling determination of the current health state of the system. Figure \ref{fig:kidm} illustrates  the KIDM architecture.

In this algorithm, we apply control inputs in two stages. First,  the encoded control inputs $\hat{K}_{i}$ are applied to update the observables. This part of the algorithms is considered as the gedradation operator in our research. Subsequently, the control inputs and observables are passed to the decoder to reconstruct the state vector. This setup,  combined  with the reconstruction loss, facilitates the separation of the degradation caused  by the applied control inputs from the  influence of the control on the state vector.

The encoder $\phi$ maps the state $x_t$ and control $u_t$ vectors to observables $y_t$. The predicted observables are concatenated with the control $u_t$ and passed to the decoder $\psi$. The reconstruction loss $\mathcal{L}_{rec}$ is used learn invertible mappings and reconstruct the state, as shown in  Fig. \ref{fig:kidm}(a). The reconstruction loss is defined as:

\begin{equation}
    \mathcal{L}_{rec} = \| x_t - \psi(\phi(x_t, u_t), u_t)\|.
\end{equation}

In order to preserve linearity of dynamics in the latent space of KIDM we use the linear dynamics loss $\mathcal{L}_{lin}$ that is given by:

\begin{equation}
    \mathcal{L}_{lin} = \| \phi(x_{t+m}) - \hat{K}_{t+m} \hat{K}_{t+m-1} ... \hat{K}_{t+1} \phi(x_t)\|,
\end{equation}

where we apply the predicted Koopman operators $\hat{K}_{i+1}$ to the observables $x_i$ over multiple time steps.

Finally, we train the model to make future state predictions Fig. \ref{fig:kidm}(b) with the future step prediction loss:
\begin{equation}
    \mathcal{L}_{pred} = \| x_{t+m} - \psi(\hat{K}_{t+m} \hat{K}_{t+m-1} ... \hat{K}_{t+1} \phi(x_t)) \|.
\end{equation}

The proposed algorithm is trained by minimizing a combination of the reconstruction error, adherence of latent dynamics to linearity and  the error of state forecasting. The total loss is given by:
\begin{equation}
    \mathcal{L} = \mathcal{L}_{rec} + \mathcal{L}_{lin} + \mathcal{L}_{pred}.
\end{equation}

\section{Case studies}
This section describes the case studies that are used to evaluate the performance of DKO and the proposed KIDM. The first subsection  provides details of  the CNC milling machine case study. In the second and third subsections, we provide  details of the Li-ion battery degradation simulation under constant and varying current load until the end of life (EoL).

\subsection{CNC milling machine}
Measuring degradation in real time is typically challenging. In  the majority of cases, the ground truth information about the health condition is  assessed only at discrete and infrequent time points. Therefore, there are only a few case studies that involve continuous monitoring of degradation.  The CNC milling machine dataset \cite{phm2010} is one of the few publicly available open source datasets that provide  continuous measurements of degradation. The dataset captures the degradation process of high-speed CNC milling machine cutters. Each cutter has three flutes that experience wear during their lifetime, and the wear of the flutes is  considered as a health parameter. The cutters are used for a total of 315 cycles under similar operating conditions. During each cycle, a set of sensors measures acoustic emission, force and three-dimensional vibration. The measurements are acquired at a frequency of 50 Hz. The dataset contains individual records of  different cutters with the ground truth health parameters provided for three cutters. The total length of recordings  ranges from $4.8 \cdot 10^7$ to $5.9 \cdot 10^7$.

The degradation of the cutter is of interest in our work due to the non-linear wear accumulation on each flute. Therefore, we can assume that this measure represents the health state and its non-linear evolution. We define the end of life of each individual cutter as the point when  the maximum flute wear reaches $150 \cdot 10^{-3}$ mm. The sensor values are collected at a high-frequency. We further process the signals to learn the dynamics in the frequency domain.

\subsection{Li-ion battery under constant current load}

For the second case study, we simulate the Li-ion battery under constant current loads using the Prognostics Models Package \cite{2022_nasa_prog_models}. This package  employs an   electrochemistry-based model to accurately simulate 18650-type cells \cite{daigle2013electrochemistry},  capturing significant electrochemical processes and aging effects.

The Li-ion battery model takes the current load as input parameter. Based on this control input, the model estimates the voltage and temperature curves. The battery model incorporates three hidden health parameters: the number of available Li-ions $q^{max}$, internal resistance $R_0$ and diffusion rate $D$.

In the first set of experiments, we simulate the battery discharge under the constant current $I_d$ of $1 A$ until the state of charge reaches 5\% of its maximum capacity. Then, the battery is charged  with a constant current of  $-1 A$ until reaching 95\% state of charge. The charge and discharge phases alternate sequentially without a rest phase. This process is repeated until reaching the end of life (EoL), which is defined as  when the battery's capacity reaches 80\% of its initial battery capacity.
We randomly sample the initial number of Li-ions and internal resistance. The number of Li-ions $q^{max}$ is drawn uniformly from the interval [7500 ions, 7600 ions], and for the internal resistance $R_0$, we draw uniformly from the interval [0.107215 Ohm, 0.127215 Ohm]. We use 70 trajectories for model training and 30 traejctories for testing.

\begin{figure}[ht]
    \centering
    \includegraphics[scale=0.5]{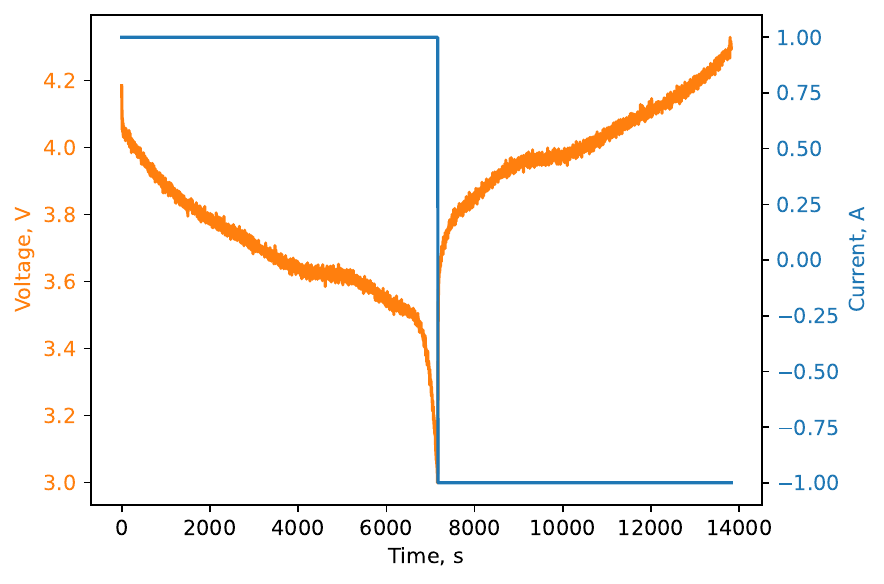}
    \caption{Sample discharge-charge cycle of the healthy Li-ion battery under constant current load. $I_d = 1 A$.}
    \label{fig:batt_const_cycle}
\end{figure}

An example of a discharge-charge cycle of the battery, simulated under a constant current load is shown in Figure \ref{fig:batt_const_cycle}.

\subsection{Li-ion battery under varying current load}

The two data cases described above serve as examples of dynamical systems without external control. In the latter case, the system undergoes a constant current load throughout its entire lifetime, rendering the impact of control negligible since it remains the same. To evaluate the performance of the KIDM algorithm, we simulate a Li-ion battery under a varying current load until it reaches the EoL \cite{2022_nasa_prog_models}. The battery is governed by piecewise constant current load, where the current magnitude and the transition point  for the next current are randomly selected. We initialize the batteries with random initial conditions, aligning them with the intervals of initial conditions used for the battery under the constant current load. 

The simulation of a fully charged battery starts with the discharge phase, where the current load follows a partially linear profile. The current load $I_d$ is  uniformly sampled from the interval [1.5A; 2.5A] . Once the current is selected, the next transition point is randomly  chosen from the interval of [100, 300] timesteps. Once the battery's state of charge reaches 0.05 of maximum charge value, a rest period of 30 timesteps begins, corresponding to 1 minute with a simulation step $dt = 2$ seconds. Afterward,  the battery is charged with a constant current of -3 A current until the state of charges reaches 0.95. The cycle then repeats until the  capacity  falls below the insufficient capacity threshold. We define  the EoL  as the point  when the battery's capacity reaches 80\% of maximum capacity. For model training we use 100 run-to-failure trajectories and 100 trajectories for testing.

\begin{figure}[ht]
    \centering
    \includegraphics[scale=0.5]{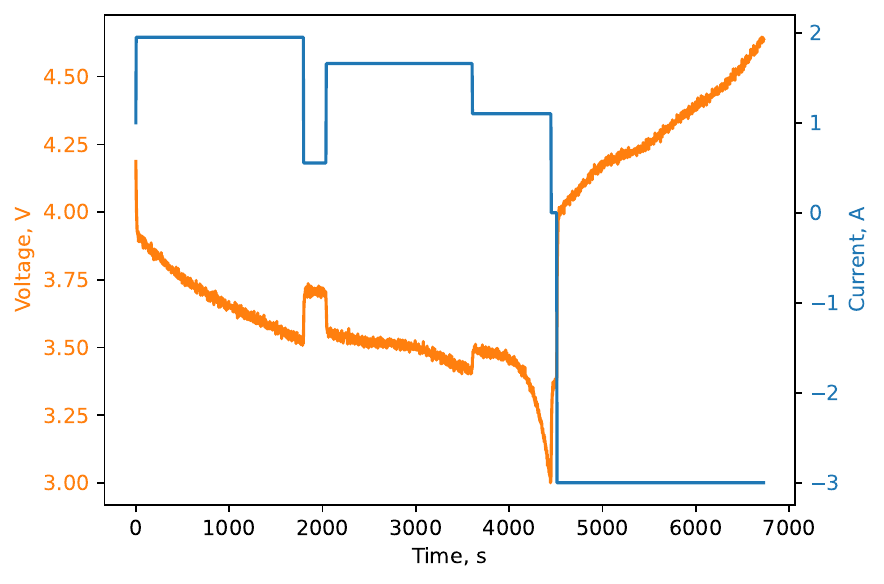}
    \caption{Sample discharge-rest-charge cycle of the healthy Li-ion battery under varying current load. $I_d \sim U(1.5A, 2.5A)$.}
    \label{fig:batt_var_cycle}
\end{figure}

Figure \ref{fig:batt_var_cycle} illustrates  the voltage and control curves for one discharge-rest-charge cycle of a healthy battery. Additionally, we  use the battery temperature curves to represent the battery state.

\begin{figure}[ht]
    \centering
    \includegraphics[scale=0.5]{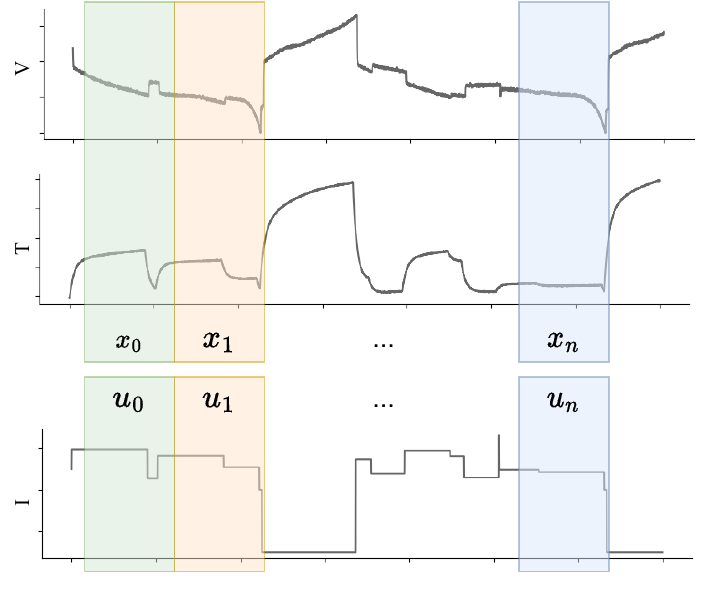}
    \caption{KIDM inputs for Li-ion battery under varying current load case.}
    \label{fig:kidm_inputs}
\end{figure}

In this work, we use the state vector $x$ and the control vector $u$ as inputs to the encoder network (see Fig. \ref{fig:kidm_inputs}). The state vector contains the voltage and battery temperature curves, while the control vector $u$ represents the current load applied to the battery.

\section{Results}

\subsection{Deep Koopman Operator}

To evaluate the performance of the DKO,  we utilize  two data cases: a CNC milling machine cutter and Li-ion battery operated under a constant current load. In the former case, we demonstrate the successful application of DKO  to high-frequency time series. Additionally, in the latter case, we showcase  the robustness of DKO against imposed noise and utilize its latent state representation to estimate the RUL.

\subsubsection{CNC milling machine cutter}

\begin{figure}[ht]
    \centering
    \includegraphics[scale=0.5]{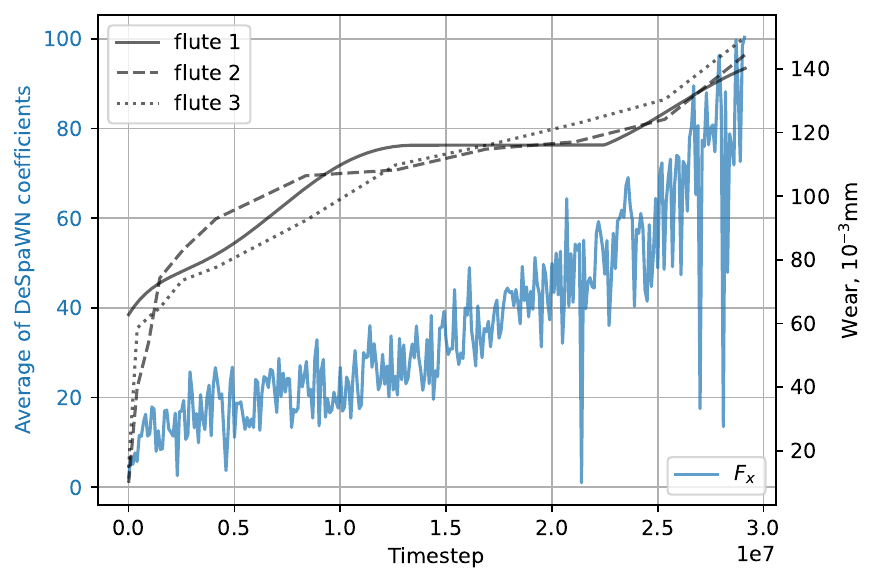}
    \caption{The average of DeSpaWN coefficients for dynamometer measurements in $x$ direction. The series are subsampled with $10^5$ stride for visualization. The measurements correspond to the cutter number six.}
    \label{fig:despawn_feature}
\end{figure}

The sensor measurements for the CNC milling machine cutters are acquired at a high frequency of 50 Hz. To train the DKO algorithm, we preprocess the initial data using the denoising sparse wavelet network (DeSpaWN) \cite{michau2022fully}. DeSpaWN applies a  cascade fast discrete wavelet transform (FDWT) with coefficient denoising. This method allows us to obtain a sparse representation of the high frequency signal. We train DeSpaWN  using eight FDWT decomposition levels and extract informative features, such as the average and maximum values of the FDWT decomposition coefficients, as well as the absolute reconstruction error of each signal.  Fig. \ref{fig:despawn_feature} displays a sample of the average DeSpaWN coefficient value for dynamometer measurements in the $x$ direction. We segment  the high frequency recording data for each cutter using a window size of 100, which corresponds to a recording of two seconds in length. This results in $2.4 \cdot 10^5$ tokens that are subsampled with a stride of$10^2$ . For model training, we use two trajectories, and one trajectory  is reserved for testing. The features of each segment are concatenated within one token and are  feed into the DKO encoder.

To learn the dynamics of the degradation process of the cutters, we use a DKO composed of a feed-forward encoder and decoder, each composed of five fully-connected layers with 100 neurons, followed by scaled exponential linear unit (SELU) activations \cite{klambauer2017self}. The dimension of the observables embedding space is set to 10. The model is optimized using the Adam algorithm \cite{kingma2014adam} with a learning rate of 0.0001 and a weight decay of $10^{-7}$ . Among the three labeled cutter datasets, two are used for training a model, while the remaining dataset is reserved  for testing. The obtained mapping to the observables space is then used to predict RUL of the cutters using  simple linear regression (LR). We compare the  performance of DKO in the RUL prediction task with an autoencoder (AE) by removing the Koopman operator part of DKO, which also serves as an ablation study. Additionally, we  compare the predictions with a feed-forward neural network (FNN) trained on the input features similar to those of the DKO encoder. The RUL predictions based on the observables of DKO demonstrate  slightly lower error, as shown in Table \ref{table:CNC_RUL}.

\begin{table}[ht]
    \centering
    \begin{center}
        \begin{tabular}{ c c c c}
            
              & MSE, $10^{-2}$ & MAE, $10^{-2}$ & MAPE, $10^{-2}$ \\ 
              \hline
            AE+LR & 2.49$\pm$1.5 & 11.03$\pm$3.22 & 42.52$\pm$17.20 \\  
            FNN & 1.94$\pm$1.21 & 9.9$\pm$2.90 & 35.95$\pm$11.43 \\
            \textbf{DKO+LR} & \textbf{1.30$\pm$1.04} & \textbf{8.52$\pm$2.94} & \textbf{30.74$\pm$1.84}
        \end{tabular}
    \end{center}
    \caption{Prediction error of the CNC milling machines RUL. The standard deviation is calculated over 5 independent initializations.}
    \label{table:CNC_RUL}
\end{table}

\begin{figure}[ht]
    \centering
    \includegraphics[scale=0.5]{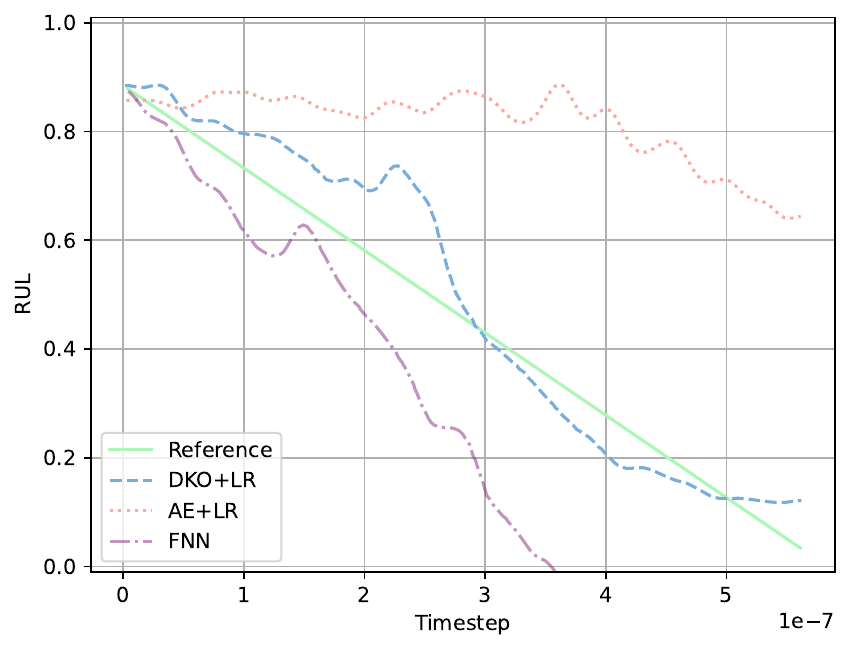}
    \caption{Sample CNC milling machie RUL prediction. Predictions are smoothed with Gaussian filter $\sigma=5$.}
    \label{fig:cnc_rul}
\end{figure}

The DKO approach has demonstrated its  capability to  accurately  determine the  degradataion trend using only  one run-to-failure trajectory, as depicted in Fig. \ref{fig:cnc_rul}. In contrast, the AE method failed to accurately  determine the RUL trend for one of the cutters. The intervals of RUL overestimation  observed in the  DKO approach correspond to periods of slow wear accumulation  and involve  dissimilar operating conditions.

\subsubsection{Li-ion battery under constant current load}

\textbf{Estimation of RUL based on learned representation.} The DKO algorithm was trained using Li-ion battery degradation trajectories under constant current load. We used100 full trajectories  for training and 100 trajectories for testing. The time series were sliced into windows of 100 points,  corresponding to a 200-seconds interval. The state space of battery dynamics was represented by voltage, temperature and current signals. The mappings $\phi$ and $\psi$ were implemented as  feed-forward networks, each consisting  of five fully-connected layers with 100 neurons followed by SELU activations \cite{klambauer2017self}. The algorithm was  optimised using the Adam algorithm \cite{kingma2014adam}  with a learning rate of 0.0001 and weight decay of $10^{-7}$. The dimensions of observables space is set to five. The linear dynamics loss $\mathcal{L}_{lin}$ and future step prediction loss $\mathcal{L}_{pred}$ for DKO were calculated over 10 consecutive time steps.

Once the dynamics of the battery are learned by DKO on several training trajectories, we use the pretrained encoder to map the battery states  to the Koopman observables space. As a result of the model construction, the dynamics in the observables space become  linear. We train LR on the Koopman observables to predict the RUL of the Li-ion battery. Since obtaining extensive data with ground truth RUL is rarely feasible  in real-life applications, we use the observables of a single full run-to-failure battery trajectory to train LR for the RUL prediction task. We compare the performance of LR model in the RUL prediction task with the LR model  on the latent space of a pretrained autoencoder (AE) model. To ensure  a fair comparison, we use the same architecture for the encoder and decoder parts of both DKO and AE.  Additionally, we evaluate the predictions of a FNN trained to predict the RUL on one run-to-failure trajectory [Table \ref{table:dko_battery}].

\begin{table}[ht]
    \centering
    \begin{center}
        \begin{tabular}{ c c c c}
            
              & MSE, $10^{-2}$ & MAE, $10^{-2}$ & MAPE, $10^{-2}$ \\
              \hline
            \textbf{AE+LR} & \textbf{0.09$\pm$0.05} & \textbf{2.50$\pm$0.84} & \textbf{25.14$\pm$9.23} \\  
            FNN & 2.09$\pm$0.48 & 12.40$\pm$1.73 & 69.53$\pm$16.49 \\
            DKO+LR & 0.13$\pm$0.04 & 2.98$\pm$0.58 & 28.07$\pm$14.81
        \end{tabular}
    \end{center}
    \caption{RUL prediction error of the Li-ion battery operated under constant current load. The standard deviation is calculated over 5 independent initializations.}
    \label{table:dko_battery}
\end{table}

\begin{figure}[ht]
    \centering
    \includegraphics[scale=0.5]{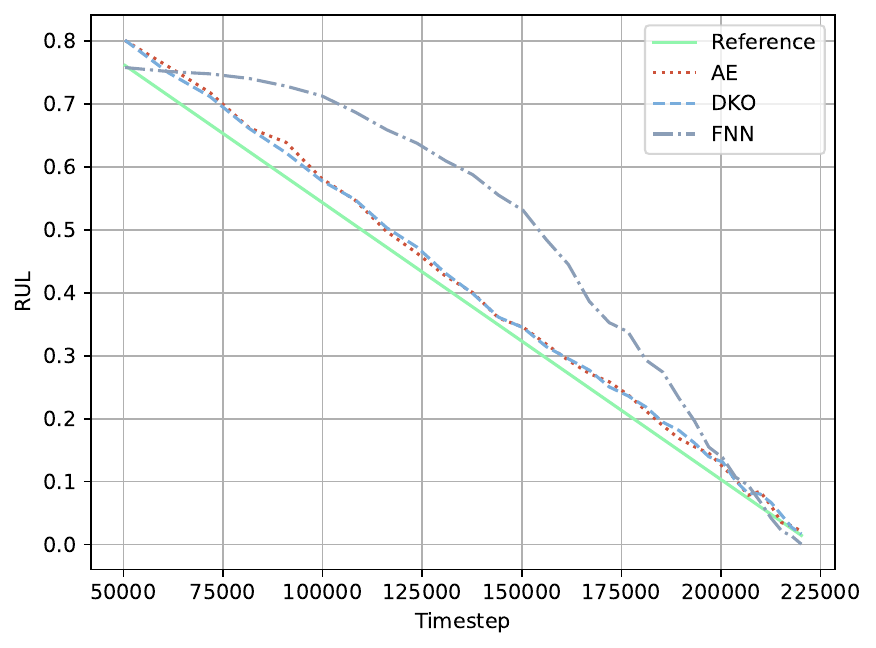}
    \caption{Sample Li-ion battery with constant current RUL prediction.}
    \label{fig:dko_battery_rul}
\end{figure}

The linear model demonstrates  comparable performance on both the Koopman observables space and the latent space of the AE. Both embeddings of the battery state provide sufficient information enough to accurately  determine the degradation trend under constant operating conditions [Fig. \ref{fig:dko_battery_rul}]. However,
the FNN fails to accurately predict the RUL due to the limited amount of available training data.

\begin{figure}[ht]
    \centering
    \includegraphics[scale=0.5]{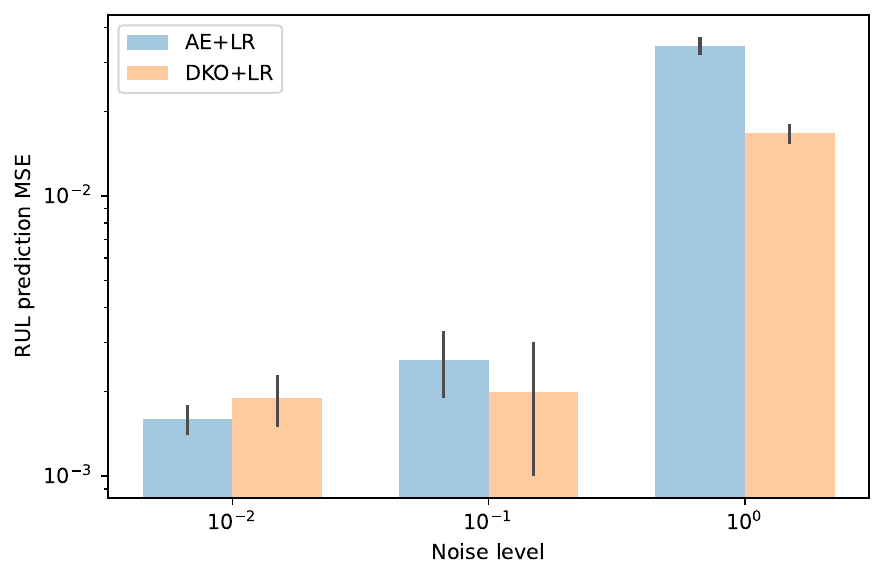}
    \caption{The RUL prediction MSE of battery under constant current load for different measurement noise levels of simulator. The shaded areas correspond to 1 standard deviation from mean.}
    \label{fig:noise_study}
\end{figure}

\noindent \textbf{Evaluation of robustness to noise.} Noise sensitivity studies provide valuable information into the reliability and stability of developed models. To evaluate the robustness to noise of DKO and the ablated models, we conducted additional experiments on simulation data with increased levels of measurement noise ranging  from 0.01 to 1.  These experiments revealed that LR trained on DKO embeddings (DKO+LR) exhibited  higher robustness  to noise and demonstrated lower RUL prediction errors compared to other models  [Table \ref{table:dko_battery_noisy}]. The results of the study are illustrated in Fig. \ref{fig:noise_study}, where the mean and standard deviation were calculated over five independent initializations.

\begin{table}[ht]
    \centering
    \begin{center}
        \begin{tabular}{ c c c c}
            & MSE, $10^{-2}$ & MAE, $10^{-2}$ & MAPE, $10^{-2}$ \\
            \hline
            AE+LR & 3.43$\pm$0.26 & 15.16$\pm$0.50 & 161.83$\pm$7.85 \\  
            FNN & 5.94$\pm$1.17 & 19.38$\pm$1.91 & 139.30$\pm$16.10 \\
            \textbf{DKO+LR} & \textbf{1.67$\pm$0.14} & \textbf{10.24$\pm$0.40} & \textbf{86.43$\pm$10.26}
        \end{tabular}
    \end{center}
    \caption{RUL prediction error of the Li-ion battery operated under constant current load with imposed measurement noise. The standard deviation is calculated over five independent initializations.}
    \label{table:dko_battery_noisy}
\end{table}

\subsection{Koopman-Ispired Degradataion Model}

\textbf{Estimation of RUL based on learned representation.} For the following experiment, we use the KIDM encoder $\phi$ and decoder $\psi$, implemented as feed-forward networks with five fully-connected layers. Each layer consists  of 100 neurons and is followed by SELU activations. The model is optimized using the Adam algorithm with a learning rate of 0.0001 and  a weight decay of $10^{-7}$. Similar to previous casy study, we set the observables dimensions to five. 

In this experiment, we train KIDM on simulation data of a Li-ion battery  operated under varying discharge current loads, denoted as $I_d \sim U(1.5A; 2.5A)$. The learned observables space is then used to train a LR model using a full run-to-failure trajectory with ground truth RUL. Furthermore, we conduct  an ablation study by removing the linear dynamics and future steps prediction part of the KIDM, resulting in an encoder-decoder model referred to as Koopman-Insipred Degradation Model's Autoencoder (KIDMAE). Additionally, we  compare the performance of the KIDM algorithm with a basic autoencoder (AE) trained on similar input features. The results, as shown in Table \ref{table:var_bat_RUL}, demonstrate the superior performance of the KIDM algorithm in the RUL prediction task with a linear model.

\begin{table}[ht]
    \centering
    \begin{center}
        \begin{tabular}{ c c c c}
            
              & MSE, $10^{-2}$ & MAE, $10^{-2}$ & MAPE, $10^{-2}$ \\
              \hline
            AE+LR & 3.19$\pm$0.12 & 14.36$\pm$0.33 & 47.41$\pm$0.63 \\
            KIDMAE+LR & 2.66$\pm$0.03 & 12.56$\pm$0.09 & 42.95$\pm$0.56 \\
            \textbf{KIDM+LR} & \textbf{0.33$\pm$0.03} & \textbf{4.65$\pm$0.18} & \textbf{13.97$\pm$0.76}
        \end{tabular}
    \end{center}
    \caption{RUL prediction error of battery operated under varying current load. The standard deviation is calculated over 5 independent initializations.}
    \label{table:var_bat_RUL}
\end{table}

\begin{figure}[ht]
    \centering
    \includegraphics[scale=0.65]{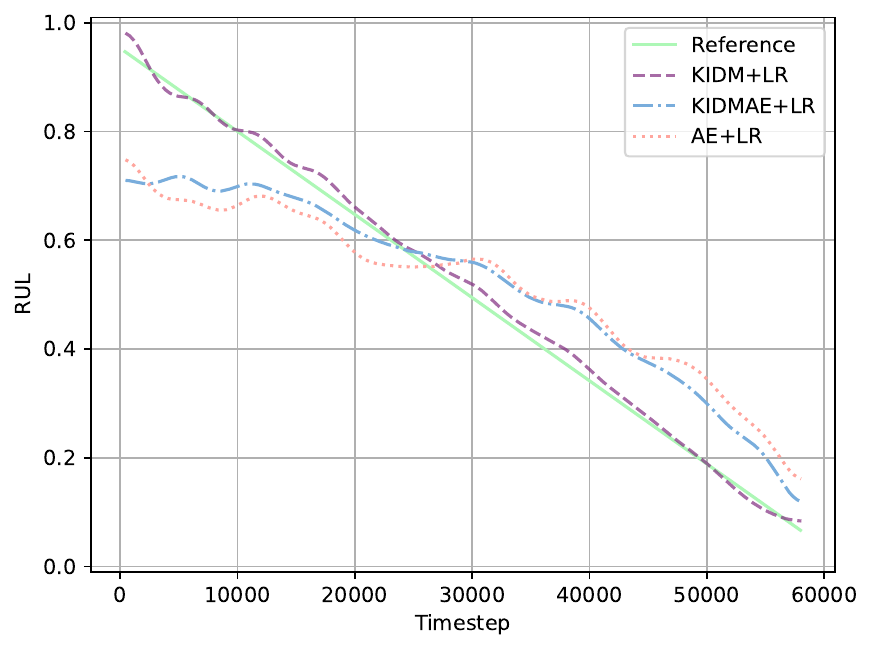}
    \caption{Sample battery under varying current RUL prediction. Predictions are smoothed with Gaussian filter $\sigma=5$.}
    \label{fig:var_batt_rul}
\end{figure}

Fig. \ref{fig:var_batt_rul} displays  a  sample RUL prediction for one of the test batteries using the KIDM algorithm.  It is evident from the figure that the algorithm successfully determined the degradation trend.

\noindent \textbf{Early lifetime RUL estimation.} In the real life industrial applications, early lifetime RUL predictions may help in optimizing maintenance strategies and enhancing overall assets reliability. In this regard, we further evaluate performance of the approach in determining the degradation trend by considering the data corresponding to the first 30\% of the battery's lifetime. The linear model is trained on this reduced data and used to predict the RUL for the next 70\% of the battery's lifetime. In this thask, KIDM demonstrates an order of magnitude lower MSE [Table \ref{table:var_bat_RUL_early}].

\begin{table}[ht]
    \centering
    \begin{center}
        \begin{tabular}{ c c c c}
            
              & MSE, $10^{-2}$ & MAE, $10^{-2}$ & MAPE, $10^{-2}$ \\
              \hline
            AE+LR & 3.47$\pm$0.78 & 15.41$\pm$1.94 & 58.18$\pm$9.96 \\
            KIDMAE+LR & 3.84$\pm$1.67 & 15.88$\pm$4.44 & 86.96$\pm$23.32 \\
            \textbf{KIDM+LR} & \textbf{0.74$\pm$0.43} & \textbf{7.54$\pm$3.13} & \textbf{35.48$\pm$14.13}
        \end{tabular}
    \end{center}
    \caption{RUL prediction error of linear model trained on first 30\% of battery lifetime. The standard deviation is calculated over five independent initializations.}
    \label{table:var_bat_RUL_early}
\end{table}

\noindent \textbf{Evaluation of extrapolation capabilities.} To test the extrapolation capabilities of the KIDM algorithm, we examine  the data of batteries simulated under different operating conditions.  Specifically, we consider different intervals of discharge currents, namely  $I \sim U(1A, 1.5A)$ and $I \sim U(2.5A, 3A)$. The results  [Table \ref{table:var_bat_extrap}] show that KIDM exhibits better performance on the data with currents drawn from the interval $[2.5A; 3A]$ . This difference in performance  may be attributed to the
 significantly slower degradation of battery health parameters under the currents drawn from the interval $[1A; 1.5A]$ .

\begin{table}[ht]
    \centering
    \begin{center}
        \begin{tabular}{ c c c c}
            & MSE, $10^{-2}$ & MAE, $10^{-2}$ & MAPE, $10^{-2}$ \\
            \hline
            & \multicolumn{3}{c}{$I \sim U(1A, 1.5A)$} \\
            \hline
            AE+LR & 15.11$\pm$2.84 & 31.4$\pm$2.52 & 108.03$\pm$10.67 \\
            KIDMAE+LR & 9.98$\pm$1.12 & 26.64$\pm$1.32 & 83.46$\pm$8.18 \\
            \textbf{KIDM+LR} & \textbf{3.12$\pm$0.58} & \textbf{14.53$\pm$1.44} & \textbf{40.34$\pm$2.58} \\
            \hline
            & \multicolumn{3}{c}{$I \sim U(2.5A, 3A)$} \\
            \hline
            AE+LR & 11.32$\pm$4.85 & 26.46$\pm$5.62 & 91.85$\pm$19.67 \\
            KIDMAE+LR & 15.31$\pm$3.37 & 30.21$\pm$3.06 & 112.07$\pm$12.03 \\
            \textbf{KIDM+LR} & \textbf{1.19$\pm$0.87} & \textbf{8.85$\pm$3.87} & \textbf{31.48$\pm$10.41} \\
            \hline
        \end{tabular}
    \end{center}
    \caption{Results of extrapolation study of RUL prediction for battery operated under varying current load. The standard deviation is calculated over 5 independent initializations.}
    \label{table:var_bat_extrap}
\end{table}

\begin{figure}[ht]
    \centering
    \includegraphics[scale=0.65]{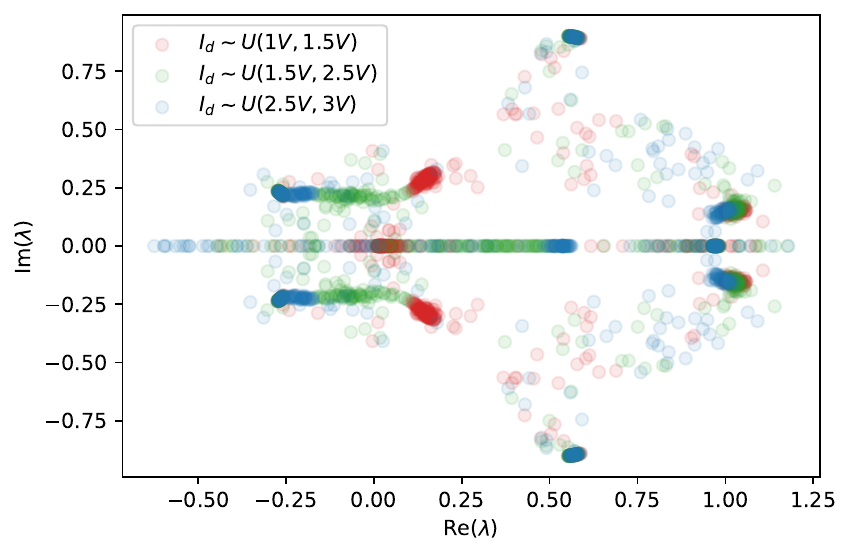}
    \caption{Eigenvalues of learned Koopman operators for different intervals of operating conditions.}
    \label{fig:koopman_eigvals}
\end{figure}

\noindent \textbf{Analysis of the latent dynamics.} To gain  a deeper understanding of the learned dynamics and the long-term behavior of the system, we analyse the predicted  Koopman operators for  batteries operated under different current ranges. Specifically, we examine their eigenvalue distributions [Fig. \ref{fig:koopman_eigvals}]. 

Each point on the plot  represents one embedding with 200 points. We  observe several clusters of complex conjugate eigenvalues, indicating the presence of fast dynamics. As expected, points corresponding to rest and charge periods overlap . However, the most interesting cluster consists of real eigenvalues distributed in the interval $Re(\lambda) \in [0; 0.6]$ . This cluster captures the long-term dynamics associated with the degradation of battery health parameters.  Notably,  we observe that lower currents impose significantly less degradation on the battery, which can explain the higher RUL prediction error for this current interval.

\section{Conclusion}

In this research, the DKO approach demonstrated its ability to learn hidden health parameters of dynamical systems without supervision. We have shown that the learned state representation can be effectively used to estimate the RUL of a system. Ablation studies indicated  that incorporating multiple steps prediction loss helps in building a more robust model that can estimate the degradation trend using a linear model for unseen CNC milling machine cutters and Li-ion batteries operated under constant current load.

Additionally, we proposed a novel approach called KIDM, inspired by DKO, to model degradation in systems under varying controls. The architecture of the KIDM model allows for minimizing the influence of controls on the observables space by separating the imposed degradataion and the controls applied to system. The KIDM model demonstrated  superior performance in RUL prediction for  Li-ion batteries operated under varying current loads. Moreover, we showed that the learned Koopman operators can be analyzed in terms of their influence on the system's health parameters.

In real-world applications,  extensive ground truth run-to-failure data is often not available. However, we have demonstrated  that both DKO and KIDM are robust in solving the RUL prediction task even with only one run-to-failure trajectory.

In this work, we have applied feed-forward neural networks to approximate the mapping to the Koopman observables space. For future research directions, more advanced architectures could be explored to improve the mapping to observables space. We believe that the approach can we successfully applied to model degradation of more complex real world dynamical systems.

\bibliographystyle{elsarticle-num} 
\bibliography{bibliography.bib}

\begin{thebibliography}{10}
\expandafter\ifx\csname url\endcsname\relax
  \def\url#1{\texttt{#1}}\fi
\expandafter\ifx\csname urlprefix\endcsname\relax\def\urlprefix{URL }\fi
\expandafter\ifx\csname href\endcsname\relax
  \def\href#1#2{#2} \def\path#1{#1}\fi

\bibitem{pimenov2023artificial}
D.~Y. Pimenov, A.~Bustillo, S.~Wojciechowski, V.~S. Sharma, M.~K. Gupta,
  M.~Kunto{\u{g}}lu, Artificial intelligence systems for tool condition
  monitoring in machining: Analysis and critical review, Journal of Intelligent
  Manufacturing 34~(5) (2023) 2079--2121.

\bibitem{coble2009fusing}
J.~COBLE, J.~Hines, Fusing data sources for optimal prognostic parameter
  selection, Transactions of the American Nuclear Society 100 (2009) 211--212.

\bibitem{daigle2012model}
M.~J. Daigle, K.~Goebel, Model-based prognostics with concurrent damage
  progression processes, IEEE Transactions on Systems, man, and cybernetics:
  systems 43~(3) (2012) 535--546.

\bibitem{daigle2016end}
M.~Daigle, C.~S. Kulkarni, End-of-discharge and end-of-life prediction in
  lithium-ion batteries with electrochemistry-based aging models, in: AIAA
  Infotech@ aerospace, 2016, p. 2132.

\bibitem{xu2021machine}
Z.~Xu, J.~H. Saleh, Machine learning for reliability engineering and safety
  applications: Review of current status and future opportunities, Reliability
  Engineering \& System Safety 211 (2021) 107530.

\bibitem{li2022end}
P.~Li, Z.~Zhang, R.~Grosu, Z.~Deng, J.~Hou, Y.~Rong, R.~Wu, An end-to-end
  neural network framework for state-of-health estimation and remaining useful
  life prediction of electric vehicle lithium batteries, Renewable and
  Sustainable Energy Reviews 156 (2022) 111843.

\bibitem{li2019deep}
X.~Li, W.~Zhang, Q.~Ding, Deep learning-based remaining useful life estimation
  of bearings using multi-scale feature extraction, Reliability engineering \&
  system safety 182 (2019) 208--218.

\bibitem{rauf2022machine}
H.~Rauf, M.~Khalid, N.~Arshad, Machine learning in state of health and
  remaining useful life estimation: Theoretical and technological development
  in battery degradation modelling, Renewable and Sustainable Energy Reviews
  156 (2022) 111903.

\bibitem{xu2022physics}
Y.~Xu, S.~Kohtz, J.~Boakye, P.~Gardoni, P.~Wang, Physics-informed machine
  learning for reliability and systems safety applications: State of the art
  and challenges, Reliability Engineering \& System Safety (2022) 108900.

\bibitem{jiang2022model}
C.~Jiang, M.~A. Vega, M.~D. Todd, Z.~Hu, Model correction and updating of a
  stochastic degradation model for failure prognostics of miter gates,
  Reliability Engineering \& System Safety 218 (2022) 108203.

\bibitem{chao2022fusing}
M.~A. Chao, C.~Kulkarni, K.~Goebel, O.~Fink, Fusing physics-based and deep
  learning models for prognostics, Reliability Engineering \& System Safety 217
  (2022) 107961.

\bibitem{koopman1931hamiltonian}
B.~O. Koopman, Hamiltonian systems and transformation in hilbert space,
  Proceedings of the National Academy of Sciences 17~(5) (1931) 315--318.

\bibitem{Lusch2017}
B.~Lusch, J.~N. Kutz, S.~L. Brunton, Deep learning for universal linear
  embeddings of nonlinear dynamics, Nature Communications 9 (2018) 4950.

\bibitem{yeung2019learning}
E.~Yeung, S.~Kundu, N.~Hodas, Learning deep neural network representations for
  koopman operators of nonlinear dynamical systems, in: 2019 American Control
  Conference (ACC), IEEE, 2019, pp. 4832--4839.

\bibitem{williams2015data}
M.~O. Williams, I.~G. Kevrekidis, C.~W. Rowley, A data--driven approximation of
  the koopman operator: Extending dynamic mode decomposition, Journal of
  Nonlinear Science 25 (2015) 1307--1346.

\bibitem{li2021deep}
M.~Li, L.~Jiang, Deep learning nonlinear multiscale dynamic problems using
  koopman operator, Journal of Computational Physics 446 (2021) 110660.

\bibitem{han2020deep}
Y.~Han, W.~Hao, U.~Vaidya, Deep learning of koopman representation for control,
  in: 2020 59th IEEE Conference on Decision and Control (CDC), IEEE, 2020, pp.
  1890--1895.

\bibitem{shi2022deep}
H.~Shi, M.~Q.-H. Meng, Deep koopman operator with control for nonlinear
  systems, IEEE Robotics and Automation Letters 7~(3) (2022) 7700--7707.

\bibitem{ping2021deep}
Z.~Ping, Z.~Yin, X.~Li, Y.~Liu, T.~Yang, Deep koopman model predictive control
  for enhancing transient stability in power grids, International Journal of
  Robust and Nonlinear Control 31~(6) (2021) 1964--1978.

\bibitem{han2021desko}
M.~Han, J.~Euler-Rolle, R.~K. Katzschmann, Desko: Stability-assured robust
  control with a deep stochastic koopman operator, in: International Conference
  on Learning Representations, 2021.

\bibitem{brunton2022modern}
S.~L. Brunton, M.~Budisic, E.~Kaiser, J.~N. Kutz, Modern koopman theory for
  dynamical systems, SIAM Review 64~(2) (2022) 229--340.

\bibitem{bevanda2021koopman}
P.~Bevanda, S.~Sosnowski, S.~Hirche, Koopman operator dynamical models:
  Learning, analysis and control, Annual Reviews in Control 52 (2021) 197--212.

\bibitem{morton2019deep}
J.~Morton, F.~D. Witherden, M.~J. Kochenderfer, Deep variational koopman
  models: inferring koopman observations for uncertainty-aware dynamics
  modeling and control, in: Proceedings of the 28th International Joint
  Conference on Artificial Intelligence, 2019, pp. 3173--3179.

\bibitem{phm2010}
X.~Li, \href{https://dx.doi.org/10.21227/jdxd-yy51}{2010 phm society conference
  data challenge} (2021).
\newblock \href {https://doi.org/10.21227/jdxd-yy51}
  {\path{doi:10.21227/jdxd-yy51}}.
\newline\urlprefix\url{https://dx.doi.org/10.21227/jdxd-yy51}

\bibitem{2022_nasa_prog_models}
C.~Teubert, M.~Corbetta, C.~Kulkarni, K.~Jarvis, M.~Daigle,
  \href{https://github.com/nasa/prog\_models}{Prognostics models python
  package} (2022).
\newline\urlprefix\url{https://github.com/nasa/prog\_models}

\bibitem{daigle2013electrochemistry}
M.~Daigle, C.~S. Kulkarni, Electrochemistry-based battery modeling for
  prognostics, in: Annual Conference of the PHM Society, Vol.~5, 2013.

\bibitem{michau2022fully}
G.~Michau, G.~Frusque, O.~Fink, Fully learnable deep wavelet transform for
  unsupervised monitoring of high-frequency time series, Proceedings of the
  National Academy of Sciences 119~(8) (2022) e2106598119.

\bibitem{klambauer2017self}
G.~Klambauer, T.~Unterthiner, A.~Mayr, S.~Hochreiter, Self-normalizing neural
  networks, Advances in neural information processing systems 30 (2017).

\bibitem{kingma2014adam}
D.~P. Kingma, J.~Ba, Adam: A method for stochastic optimization, arXiv preprint
  arXiv:1412.6980 (2014).

\end{thebibliography}
 

\end{document}